# Different approaches to generate matching effects using arrays in contact with superconducting films.

**J del Valle[1,a)], A Gomez[1,b)], J L Hita[1,c)], V Rollano[1], E M Gonzalez[1,2] and J L Vicent[1,2]**

[1]Departamento Física de Materiales, Facultad CC. Físicas, Universidad Complutense, 28040 Madrid, Spain.
[2]IMDEA-Nanociencia, Cantoblanco, 28049 Madrid, Spain.

E-mail: jlvicent@ucm.es

**Abstract.** Superconducting films in contact with non-superconducting regular arrays can exhibit commensurability effects between the vortex lattice and the unit cell of the pinning array. These matching effects yield a slowdown of the vortex flow and the corresponding dissipation decrease. The superconducting samples are Nb films grown on Si substrates. We have studied these matching effects with the array on top, embedded or threading the Nb superconducting films and using different materials (Si, Cu, Ni, Py dots and dots fabricated with Co/Pd multilayers). These hybrids allow studying the contribution of different pinning potentials to the matching effects. The main findings are: i) Periodic roughness induced in the superconducting film is enough to generate resistivity minima; ii) A minor effect is achieved by magnetic pinning from periodic magnetic field potentials obtained by dots with out of plane magnetization grown on top of the superconducting film, iii) In the case of array of magnetic dots embedded in the films vortex flow probes the magnetic state; i.e. magnetoresistance measurements detect the magnetic state of very small nanomagnets. In addition, we have studied the role played by the local order in the commensurability effects. This was attained using an array that mimics a smectic crystal. We have found that preserving the local order is crucial. If the local order is not retained the magnetoresistance minima vanish.

*Pacs numbers:* 74.25.F-, 74.25.Wx, 74.78. Na, 74.25.Ha

------------------------------------------

a) Present address: Department of Physics, University of California-San Diego, CA 92093, USA.

b) Present address: Centro de Astrobiologia, INTA-CSIC, 28850 Torrejón de Ardoz, Spain.

c) Present address: Departamento Fisica Materia Condensada, Universidad Autonoma de Madrid, 28049 Madrid, Spain.

# 1. Introduction

Vortex pinning [1, 2] is one of the main topics in applied superconductivity. Pinning enhancement has been achieved using different strategies [3-5]. The problem of vortex pinning is present in superconductivity large scale applications as well as superconducting based devices. Superconducting vortices play distinct roles in many devices. They are the needed objects in several superconducting devices, as for example superconducting related rectifier devices [6, 7] or superconducting related memory devices [8, 9]. On the other hand, vortices are the source of noise [10] in SQUID devices. Many different approaches have been used to pin and control vortices. One of the most promising approaches is based on fabricating the appropriate nanostructures in the superconducting sample. This technique has been a successful method to achieve a noticeable control and increase of vortex pinning. Samples with modulated thickness [11] or samples with alternately superconducting and non-superconducting layers [12, 13] are relevant examples. The recent progress in nanofabrication techniques [14] has opened new avenues and many different types of ordered pinning centers have been designed. One of the most significant achievements is the commensurability effects between the vortex lattice and the ordered defects [15, 16]. When this condition is fulfilled a drastic decrease of dissipation occurs for selected values of the applied magnetic field; i. e. an increase in critical current happens.

In the reported experiments, see reference [17] and references therein, several pinning mechanisms usually act simultaneously; for instance, the periodic roughness induced in the film in the case of embedded arrays of defects, or the adjustable strength of stray magnetic fields in the case of arrays of nanomagnets [18], etc. In the pioneering paper of Baert el al. [15], the array of defects is made using the so-called antidots; i. e. array of holes in the superconducting film. In our present study we only focus on dots which allow using different materials, so they can generate different pinning potentials. Different shapes of pinning centers have been reported in the literature [19-23] for examples lines, rings, squares, etc. In this work, we use only arrays of circular dots. Hence, we can focus on matching effects induced by artificial periodic pinning potentials where the origin of the commensurability effect is not the pinning center shape.

Another relevant topic is the array geometry. Several works have dealt with different order/disorder arrays, as well as arrays with peculiar symmetry configurations [25-32]. An enhancement of pinning forces can be obtained with array of defects which present a non-uniform density, but keeping the local ordering, or with a graded distribution of defects. In both cases matching effects are obtained. An open question is whether or not local order is a needed condition to observe matching effects.

In order to shed light on these subjects, we study the effect of pinning centers fabricated with different materials which sculpt different pinning potentials and we explore the effect of breaking the local order of pinning potentials.

To accomplish these goals we have fabricated superconducting films with controlled roughness, with magnetic and non- magnetic dots, with dots embedded in the superconducting films, with dots on top of the superconducting films and with dots threading the superconducting films.

We have observed such a rich scenario that allows us tailoring matching effects with different outcomes. Finally, by studying the effect of pinning order we observe that the lack of order in one direction of the pinning array is enough to prevent matching effects.

The paper is organized as follows. First, we present the types of hybrid samples and the fabrication and measurement techniques. Then, we present the results in superconducting film with i) induced periodic roughness;  ii)  superconducting films with array of magnetic dots grown on top of the film to study the effect of periodic roughness and periodic stray magnetic fields respectively. Then, we present the results obtained in samples with embedded array of magnetic and non-magnetic dots to study the competition between them. Then, hybrid sample fabricated with magnetic dots which thread the sample. Pinning hysteresis effects can be observed in this type of hybrid sample. Finally, we present the results obtained in a sample with an array that mimics a smectic crystal. This hybrid sample allows us exploring whether or not preserving the local order of pinning centers is crucial to obtain commensurability effects.

## 2. Experimental details

We have fabricated superconducting/magnetic hybrids on Si (100) substrates by sputtering and electron beam lithography techniques. The pinning arrays are based on circular nanodots. These nanodots are made with different materials: Si, Cu, Ni, Py and Co/Pd multilayers. The nanodot height is always 40 nm and the dot diameter is 200 nm. The only exception is the hybrid sample whose dots are threading the film. In this case the dot height is 160 nm and diameter is 180 nm.

The fabrication technique follows the usual methods. First, we performed electron beam lithography over a Si substrate coated with PMMA. Afterwards, the material of the non-superconducting pinning centers was sputtered on top of the resist, and then we lifted off to define the nanodot array. After this, a superconducting 100 nm Nb film was grown using magnetron sputtering. In the case of the dots threading the Nb, the height of the Ni dots was 160 nm, which ensures the dots will completely perforate the Nb film. The only exception to this general fabrication process is the Co/Pd dot sample: the dots were deposited on top of the array, so the lithography and the multilayer growth were done on top of the Nb film. The dots are arranged on regular lattices (square or rectangular); only to address the influence of directional disorder we have fabricated a hybrid sample whose array mimics a smectic crystal. Smectic order is characterized for keeping translational periodicity in one direction while order is lost in the other directions. In our particular 2D case, a smectic array would consist in rows of Cu nanodots, but dot position is randomly distributed within each row. We have patterned such dot configuration by means of e-beam lithography by defining each dot position ($x_{i,j}$, $y_{ij}$) in the following way: $x_{i,j} = i\,a$; $y_{i,j} = j\,a + \alpha$. Where $i$ and $j$ are integers and $\alpha$ is a random number from $-a/2$ to $+a/2$, being $a = 490$ nm the lattice parameter.

Conventional lithography and etching techniques allow patterning an appropriate cross-shaped bridge (40 x 40 $\mu m^2$ area) for transport measurements. Magnetotransport is measured using a commercial Helium cryostat with variable temperature insert and a superconducting solenoid. The magnetic field is applied normal to the sample plane. Vortices are driven by dc currents injected in the patterned cross-shaped bridge. The standard four point method was used for the measurements; data are taken close to critical temperature. In each hybrid system the current density used is the optimum current density for the observation of periodic pinning, see reference 24.

## 3. Results and discussion

First, we have studied a sample which is made up for an array with 400 nm x 400 nm unit cell of Si dots (40 nm height and 200 nm diameter) and a 100 nm thick Nb film deposited on top of the array. That is, the Si dots, grown on top of the Si substrate, result in a wrinkled substrate with a regular distribution of *nanohills*. Magnetoresistance data are shown in figure 1. A strong matching effect is obtained when the applied magnetic field fulfills the matching condition. The drop in resistivity is more than one order of magnitude.

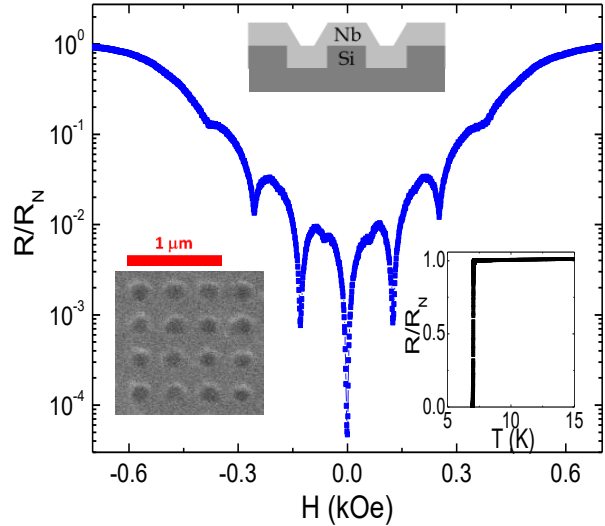

**Figure 1.** Normalized resistance vs. applied magnetic fields. Sample Nb film (100 nm thickness) with embedded square array (400 x 400 nm unit cell) of Si dots (40 nm thickness, 200 nm diameter) T= 0.99 $T_c$. J=3·10$^8$ A/m$^2$. Upper inset shows hybrid sample sketch. Left inset shows SEM image of the array. Right inset shows normalized resistance vs. temperature ($R_N$ normal state resistance). Colour on line.

The first minimum appears at magnetic field B = ($\Phi_0$/A), where A is the area of the array unit cell and $\Phi_0$= 20.7 G μm$^2$ is the quantum fluxoid. More minima can appear at matching fields $B_n$ = n ($\Phi_0$/A), where n >1 is an integer number. Under these conditions, the vortex lattice and the pinning array are commensurate. Sometimes minima can be also observed at fractional matching fields $H_f$ = f ($\Phi_0$/A), being f a non-integer number. In summary, the vortex lattice motion slows down when the matching conditions are fulfilled and consequently a strong diminishing in dissipation is observed.

Next, we have studied the influence of magnetism in vortex pinning. Magnetism and superconductivity are competing effects. However, in recent years, magnetism has been used to model, enhance and modify superconducting properties [33-38]. Concerning matching effects, the most remarkable outcome is a noticeable asymmetry in critical current maxima when the applied magnetic fields are reversed [33]. This happens when the dots show a strong out of plane magnetization. Another striking result is the so-called magnetic-field-induced superconductivity [34]. Lange et al. [34] place the magnetic array on top of the superconducting film. These authors claim a selective enhance of the superconducting critical fields. This increase in superconductivity is induced by compensation of the stray fields coming from dot magnetization by external applied field among dots. We have fabricated with the same ingredients and configuration a hybrid sample similar to that reported by Lange et al. [34]. The hybrid sample has been made of dots of Co/Pd multilayers arranged in a 400 x 400 nm array unit cell on top of the Nb film. The multilayer period has been chosen to produce magnetic dots with out of plane magnetization. The fabrication and magnetic characterization details can be found in reference 37. In that work the Co/Pd multilayers were embedded in the Nb film and magnetoresistance minima were observed. In the present work the array and dot dimensions are the same that in the previous hybrid sample with Si dots. Fig. 2 shows the magnetoresistance data.

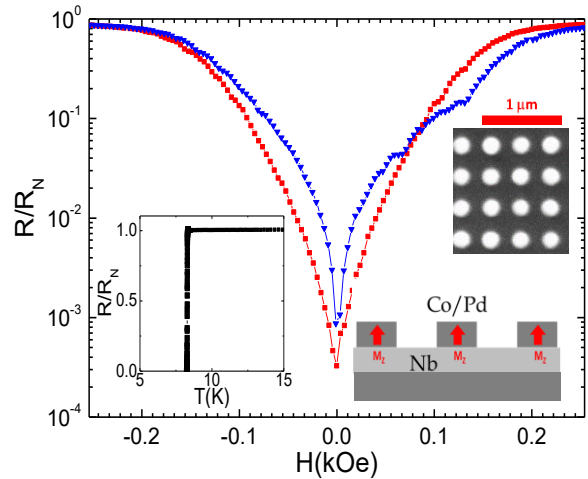

**Figure 2.** Normalized resistance vs. applied magnetic fields. Sample Nb film (100 nm thick) with on top square array (400 x 400 nm unit cell) of Co/Pd dots (0.4 nm Co layer ; 0.6 nm Pd layer ; total thickness of the multilayer 40 nm, dot diameter 200 nm). T= 0.99 $T_c$. J=0.5·10$^8$ A/m$^2$. Triangles: upwards out of plane remanent magnetization of the Co/Pd array. Squares: the Co/Pd array is in the demagnetized state. Upper right inset shows SEM image of the array. Lower right inset shows hybrid sample sketch. Left inset shows normalized resistance vs. temperature ($R_N$ normal state resistance). (Colour on line).

The matching effect is very weak, almost vanishes in comparison with the sharp minima induced by periodic roughness. We observe a slightly asymmetry in the magnetoresistance. This is the subject that we are going to address next.

This magnetoresistance asymmetry effect is related to out of plane magnetization in the magnetic dots. This effect is the previously mentioned asymmetry in the pinning effect when the applied magnetic fields are reversed [33]. We have fabricated a similar sample to the one that was reported by Morgan and Ketterson [33]; i. e. array of Ni dots which thread the superconducting Nb film. In our configuration the array is 400 x 400 nm array unit cell, the height of the Ni dot is 160 nm and the Ni dot diameter is 180 nm. Figure 3 shows the usual magnetoresistance data showing the already reported asymmetry in the minima [33].

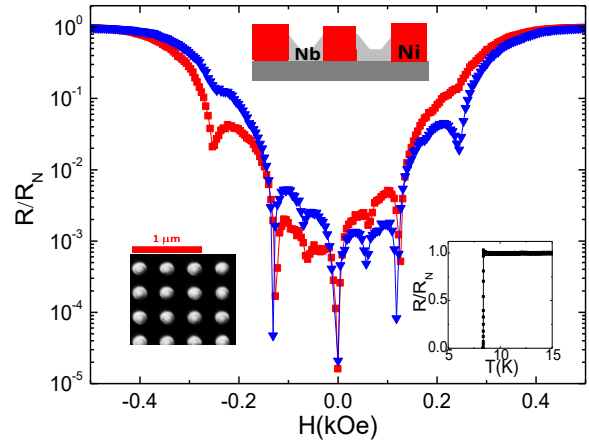

**Figure 3.** Normalized resistance vs. applied magnetic fields. Sample Nb film (thickness 100 nm) with square array (400 x 400 nm unit cell) of Ni dots (thickness 160 nm and diameter 180 nm). T= 0.99 $T_c$. Triangles: upwards out of plane remanent magnetization of the Ni array. Squares: downwards out of plane remanent magnetization of the Ni array. Upper inset shows hybrid sample sketch. SEM image of the array. Lower left inset shows SEM image of the array. Lower right inset shows normalized resistance vs. temperature ($R_N$ normal state resistance). (Colour on line).

Moreover, fractional minima are obtained when the applied magnetic field is ½ $H_1$, $H_1$ being the first matching field. In this condition and for all magnetic fields applied below the first matching field value, all the vortices are trapped vortices. The first matching field separates a state with all the vortices trapped from another state in which trapped vortices coexist with interstitial vortices.

In figure 4 we focus on applied magnetic fields around this fractional matching field, when all the vortices are trapped. We have measured the dissipation responses when the Ni dots display three characteristic magnetic states: i) demagnetized state, which is reached, at usual, with vanishing minor hysteresis loops obtained by applying decreasing magnetic fields; ii) upwards remanent magnetization and iii) downwards remanent magnetization, which are obtained by applying a 30 kOe saturating magnetic fields in the appropriate direction and then switching them off. As can be seen, the magnetoresistance data at fractional matching fields change with the magnetic state of the Ni dot. We can obtain the magnetic state of very tiny nanomagnets by detecting the motion of superconducting vortices. In summary, the matching effect is mostly due to periodic roughness in the superconducting film, but an appropriate design of magnetic pinning potentials can subtlety tailor the interplay between superconducting and magnetic properties at the matching fields. We can underline that vortex motion probes very efficiently the magnetic state of tiny nanomagnets.

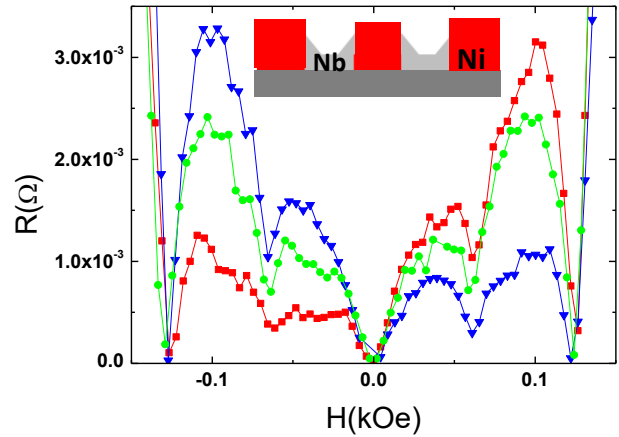

**Figure 4.** Resistance vs. applied magnetic fields around fractional ½ $H_1$ minima ($H_1$ being the first minimum). Sample Nb film (100 nm thick) with square array (400 x 400 nm unit cell) of Ni dots (160 nm thick, 180 nm diameter). T= 0.99 $T_c$. Triangles: dots with upwards out of plane remanent magnetization. Squares: dots with downwards out of plane remanent magnetization. Circles: the dot in the demagnetized state. Inset shows hybrid sample sketch. (Colour on line).

To have a complete overview regarding commensurability and pinning, we have to study the effect of embedded magnetic dots without out of plane magnetization. So, we have to compare matching effects induced in the superconducting film by embedded magnetic array and embedded non-magnetic array. We have fabricated two samples keeping the same array pattern arrangement, one sample with Cu dots and the other one with Py dots. The array unit cell in both samples is 400 nm x 600 nm. The Py magnetic dots are in a magnetic vortex state with a small vortex core [39]. Figure 5 shows the same matching effect in both cases. The only outcome is related to the different background dissipation, this effect is a signature of the magnetic state of the dots as was reported recently [37]. In summary, the periodic roughness is enough to yield commensurability effects, and the interplay between stray fields and ordered magnetic pinning potentials are a very good tool to tailor the dissipation response and to probe the magnetic state of tiny nanomagnets.

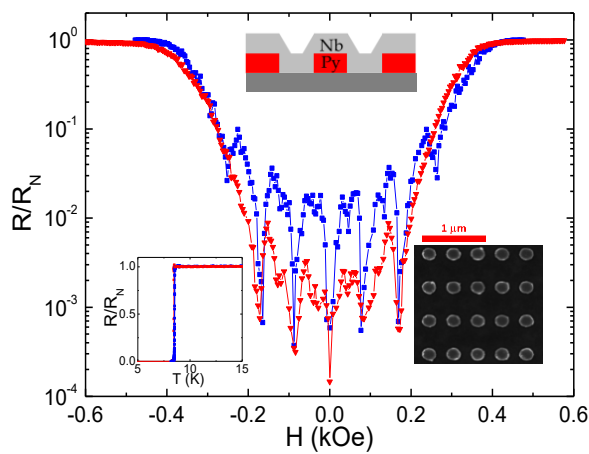

**Figure 5.** Normalized resistance vs. applied magnetic fields. Sample Nb film (100 nm thickness) with dot array, 400 nm x 600 nm unit cell and 40 nm thickness and 200 nm diameter dots. Triangles: Py dot sample, magnetic state with aligned polarities of the magnetic vortex cores. Squares: Cu dot sample. T= 0.99 $T_c$. Upper inset shows sample sketch. Lower left inset shows normalized resistance vs. temperature ($R_N$ normal state resistance); red triangles: Py dot sample; blue squares: Cu dot sample. Lower right inset SEM image of the Cu dot array. (Colour on line)

Finally, we explore some of the constraints that the array has to fulfill to generate matching effects. Recent works have reported [25-32] an enhancement of pinning effects when the possible channels for vortex motion are precluded, for example by dots arranged in a conformal crystals. This pinning increase has been realized with array of pinning centers which are designed with non-uniform density while preserving the local ordering of the pinning potentials. In our case, we study commensurability effects when the pinning potential local order is not conserved. Vortex matter is a very peculiar state of matter and it is well known that vortex lattice moving on ordered [40] as well as random pinning potentials [41] can trigger a vortex smectic phase. Surprisingly, perfect regular arrays can induce both vortex commensurability effects and smectic phase [40]. So, arrays which mimic a smectic crystal seem to be a right choice to study what happens if we disturb the local order in such a way that we uphold the order periodicity in one direction and we break the periodicity in the perpendicular direction. We know that vortex lattices can adopt smectic order easily. Therefore, at least, we would expect matching effect in one of the direction of vortex motion; that is, in the direction where the pinning potentials show translational periodicity. To study this, we have grown a sample with Cu dots embedded in Nb film arranged in a smectic–like shape. That is, the Cu dots only show translational periodicity in one direction. Figure 6 shows the magnetotransport data in both directions, parallel and perpendicular to the direction in which the dots are ordered in rows. First of all, figure 6 shows strong direction dependence of the magnetoresistance data. This behavior is similar to the one already reported in reference 19. In that case the hybrid sample is Ni lines embedded in Nb film. Both samples exhibit different behaviors when the current is applied parallel to the channels among the periodic lines or rows of dots and when the current is applied perpendicular to the periodic defects (lines or rows of dots).

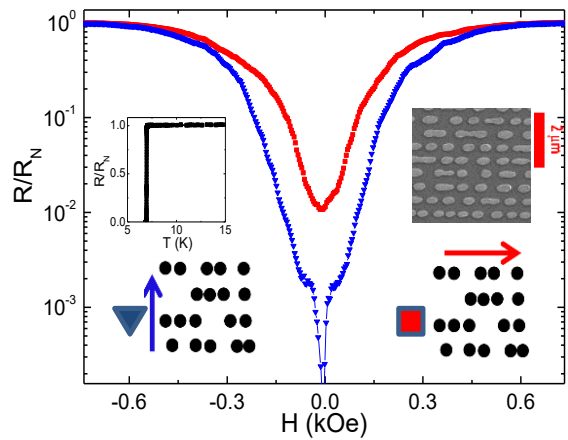

**Figure 6.** Normalized resistance vs. applied magnetic fields. Hybrid sample array of Cu nanodot mimicking a smectic crystal embedded in a 100 nm Nb film. Periodicity in the y-axis

direction is 490 nm (see text) of Cu dots (40 nm thick, diameter 200 nm). T= 0.99 $T_c$. Lower insets show array sketch and the direction of vortex motion. Triangles show magnetoresistance with vortex motion parallel to the direction of ordered pinning potentials. Squares show magnetoresistance with vortex motion parallel to the random pinning potential direction. Upper left inset shows normalized resistance vs. temperature ($R_N$ normal state resistance). Upper right inset SEM image of the Cu dot array. (Colour on line).

Interestingly, the lack of local order in one direction (smectic-like array) annihilates the matching effects utterly. We can conclude that when pinning potentials do not preserve the local order, the commensurability effects vanish. This result can be understood following the analysis and findings of the vortex-nanodot interaction reported in [40]. According to this work the coherent length is the crucial parameter. Figure 7(a) shows the vortex-nanodot interaction range for a square array and for a smectic-like array in two different situations. First, at low temperature, i. e. in the temperature region where the sample intrinsic pinning overcomes the artificially induced pinning potentials, the lattice vortex dynamics do not show matching effects, see [42]. We have chosen temperature $0.4T_c$ which corresponds to coherence length 0.1 times the array lattice constant. Then, we have estimated the vortex-pinning interaction for temperature $0.99\ T_c$ which corresponds to coherence length 0.5 times the array lattice constant. At this temperature the lattice vortex dynamics commensurability effects should be crystal clear since, close to $T_c$, the ordered pinning potentials overcome the intrinsic random potentials. But the lattice vortex dynamics for the smectic sample deviates from this scenario, since commensurability effects are always absent. To work out this behavior we have studied the Fourier transforms of the previous interactions, see figure 7(b). An interesting effect can be pointed out: while for a regular square array the symmetry is preserved even for temperatures pretty close to $T_c$; for the smectic case, order is lost also along the ordered direction as the range of the interaction becomes larger (higher temperature). The stripe-like Fourier transform evolves into an isotropic pattern. This analysis explains the results observed in figure 6. Breaking the order in one direction does also diminish it along the other direction, and the commensurability effect vanishes.

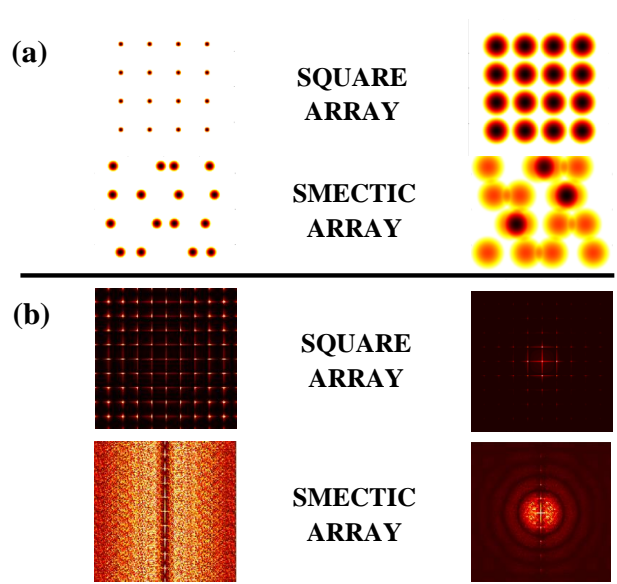

**Figure 7.** (a) Left: vortex- nanodot interaction at 0.4 $T_c$ for square and smectic-like arrays. Right: vortex- nanodot interaction at 0.99 $T_c$ for square and smectic-like arrays. (b) Left: Fourier transform of the square and smectic-like arrays at 0.4 $T_c$. Right: Fourier transform of the square and smectic-like arrays at 0.99 $T_c$. (Colour on line).

## 4. Conclusions

In summary, many superconducting electronic devices rely on the control of vortex flow and pinning of vortex lattice. Regular arrays of defects is a powerful tool to work out this challenge, for instance this approach has been used in devices as different from one another as superconducting microwave resonators [43] or superconducting rectifier [6]. In this work, we have studied the needed conditions to commensurate the vortex lattice with the array of pinning potentials. We observe that the sharp decrease in dissipation at matching conditions can be obtained with periodic roughness and nothing else. The combination of structural periodicity with stray magnetic fields can yield mechanisms to control the vortex flow and to probe the magnetic state of nanomagnets. Once the matching effect is generated in the case of magnetic dots with strong magnetic stray fields, vortex lattice motion can discriminate the magnetic state of very tiny nanomagnets. Finally, matching effects show up only when perfect local order around the pinning potentials exists. Only when the vortex lattice and the ordered array meet the commensurability conditions the vortex flow slows down and a noticeable diminishing in the sample dissipation emerges.

Acknowledgments

We thank support from Spanish Ministerio de Economia y Competitividad grant FIS2013-45469 and CAM grant P2013/MIT-2850 and EU COST Action MP1201.